\documentclass[12pt,a4paper]{article}
\usepackage{graphicx}
\usepackage{times}
\def\aj{AJ}%
\def\araa{ARA\&A}%
\def\apj{ApJ}%
\def\apjl{ApJ}%
\def\aap{A\&A}%
\def\aaps{A\&AS}%
\def\mnras{MNRAS}%
\def\zap{ZAp}%
\def\nat{Nature}%
\title{\bf Red Supergiants, Luminous Blue Variables and Wolf-Rayet stars: the single massive star 
perspective}
\author{Georges Meynet$^1$, Cyril Georgy$^1$, Raphael Hirschi$^2$, Andr\'e Maeder$^1$,\\
Phil Massey$^3$, Norbert Przybilla$^4$, M.-Fernanda Nieva$^5$\\
\normalsize $^1$ Geneva Observatory, University of Geneva, Maillettes 51, 1290 Sauverny, Switzerland\\ 
\normalsize $^2$ Astrophysics Group, EPSAM Institute, University of Keele, Keele, ST5 5BG, UK\\
\normalsize $^3$ Lowell Observatory, 1400 W Mars Hill Road, Flagstaff, AZ 86001, USA \\
\normalsize $^4$ Dr. Karl Remeis-Observatory \& ECAP, Sternwartstr. 7, 96049 Bamberg, Germany\\
\normalsize $^5$ Max-Planck-Institut fur Astrophysik, Karl-Schwarzschild-Str. 1, 85741 Garching, Germany\\
}
\date{\mbox{}}
\begin{document}
\maketitle
\pagestyle{empty}
%
%
\def\bull{\vrule height .9ex width .8ex depth -.1ex}
\makeatletter
\def\ps@plain{\let\@mkboth\gobbletwo
\def\@oddhead{}\def\@oddfoot{\hfil\tiny\bull\quad
``The multi-wavelength view of hot, massive stars''; 39$^{\rm th}$ Li\`ege Int.\ Astroph.\ Coll., 12-16 July 2010 \quad\bull}%
\def\@evenhead{}\let\@evenfoot\@oddfoot}
\makeatother
%
%
\def\beginrefer{\section*{References}%
\begin{quotation}\mbox{}\par}
\def\refer#1\par{{\setlength{\parindent}{-\leftmargin}\indent#1\par}}
\def\endrefer{\end{quotation}}
%
%
{\noindent\small{\bf Abstract:} 
We discuss, in the context of the single star scenario, the nature of the progenitors of Red Supergiants (RSG), of Luminous Blue Variables (LBV) and of Wolf-Rayet (WR) stars. These three different populations correspond to evolved phases of Main Sequence (MS) OB stars. Axial rotation and mass loss have a great influence on massive star evolution in general and more specifically  on the durations of these different phases. Moderate rotation and mass loss, during the MS phase, favor the evolution towards the RSG stage. Fast rotation and strong mass loss during the MS phase, in contrast, prevent the star from becoming a RSG and allow the star to pass directly from the OB star phase into the WR phase. Mass loss during the RSG stage may make the star evolve back in the blue part of the HR diagram. 
We argue that such an evolution may be more common than presently accounted for in stellar models. 
This might be the reason for the lack of
type IIP SNe with RSG progenitors having initial masses between 18 and 30 M$_\odot$.
The LBVs do appear as a
possible transition phase between O and WR stars or between WNL and WNE stars. Fast rotation and/or strong mass loss during the Main-Sequence phase prevent the formation
of LBV stars.
The mechanisms driving the very strong ejections shown by LBV stars are still unknown. We present some
arguments showing that axial rotation together with the proximity of the Eddington limit may play a role in driving the shell ejections. 
Rotation and mass loss favor the formation of Wolf-Rayet stars. 
The fact that WR stars and RSGs rarely occur in 
the same coeval populations indicates that the mass range of these two populations is different, WR stars originating from more massive stars than RSGs.
Single star evolution models predict variations with the metallicity of the number ratios of Type Ibc to Type II supernovae, of Type Ib to Type II and of Type Ic to Type II,
which are compatible with observations,
provided that many stars leaving a black hole as a remnant produce an observable supernova event.}
%
%
\section {Introduction}

Red Supergiants (RSG), Luminous Blue Variables (LBV) and Wolf-Rayet (WR) stars represent phases during which the star may lose large amounts of mass by stellar winds
and/or strong shell ejections. These mass loss events are on the whole still not well understood. Our ignorance of their driving mechanisms, amplitudes, durations, metallicity dependence is in great part responsible for the uncertainties pertaining the results of stellar models. Very schematically, 
the evolutionary scenarios, at solar metallicity, for single massive stars follow the sequences shown below:
\vskip 5mm
\noindent
{\bf \underline{{$M >90  M_{\odot}$}}}:  O –- Of –- WNL –- (WNE) -– WCL –- WCE -– 
SN(SNIbc/BH/SNIIn)?  (PCSN/Hypernova low Z?)\\
\noindent
{\bf \underline{{$60-90 \; M_{\odot}$}}}: O –- Of/WNL$\Leftrightarrow$LBV -– WNL(H poor)-– WCL-E -– SN(SNIbc/BH/SNIIn)?\\
\noindent
{\bf \underline{{$40-60 \; M_{\odot}$}}}: O –- BSG –-  LBV $\Leftrightarrow$ WNL -–(WNE) -- WCL-E –- SN(SNIb) \\
\hspace*{5.9cm}  - WCL-E - WO – SN (SNIc) \\
\noindent
{\bf \underline{{$30-40 \; M_{\odot}$}}}:  O –- BSG –- RSG  --  WNE –- WCE -– SN(SNIb)\\
\hspace*{4.0cm}                        OH/IR $\Leftrightarrow$ LBV ? \\
\noindent
{\bf \underline{{$20-30 \; M_{\odot}$}}}: O -–(BSG)–-  RSG  -- BSG (blue loop) -- RSG  -- SN(SNIIb, SNIIL)\\
\noindent
{\bf \underline{{$10-20 \; M_{\odot}$}}}: O –-  RSG -– (Cepheid loop, $M < 15 \; M_{\odot}$) – RSG -- 
SN (SNIIP)\\ 

The sign $\Leftrightarrow$ means back and forth motions between the two  stages. The various types of supernovae are tentatively indicated. The limits 
between the various scenarios  depend on metallicity $Z$ and rotation. 

We can classify the different scenarios in three families: below about 30 M$_\odot$, stars go through a red supergiant stage (RSG); above about 40 M$_\odot$,
stars go through a Luminous Blue Variable (LBV) and a Wolf-Rayet (WR) phase\footnote{WN shows at their surface the products of H-burning, WC and WO, the products
of He-burning, see more details in Crowther (2007).}; in the intermediate mass range, between  30 and 40 M$_\odot$,  stars can possibly (but not necessarily) go through the three phases, RSG, LBV and WR. As we shall discuss below, many very interesting questions remain to be solved before we can safely say that we
understand massive star evolution. Among the challenges, which still remain before us, let us cite:
\begin{itemize}
\item The finding of a  consistent explanation of the overall dependence with metallicity of the blue to red (B/R) supergiant ratio (see more below).
\item The driving mechanism for the huge shell ejections undergone by LBV stars.
\item The importance of the single and binary channels for explaining the WR and the different core collapse supernova types (II, Ib, Ic) at various metallicities.
\end{itemize}
In the following we discuss the importance of mass loss and rotation in the context of these questions and we propose some directions of research for future investigations.
Very interesting and informative reviews on these topics have recently be written by Massey (2003), Crowther (2007) and Smartt (2009).

\section{Red Supergiants: impact of mass loss and rotation}

For a long time there was a mismatch between the observed positions of RSGs in the HR diagram and the predicted ones.
Improvements in model atmospheres and in interior models allow now to obtain a good agreement, in particular Levesque et al. (2005)
find that the upper luminosity of RSGs is around 5.2-5.3 (in log L/L$_\odot$ and taking the median value of the five most luminous RSGs).
Interestingly, this upper luminosity presents no variation with the metallicity (Levesque et al. 2006). Red supergiants are found to be variable in the
V band (not in K) with amplitude of 0.5 mag. in M31 and of 0.9 mag. in the Magellanic Clouds (Massey et al. 2009). Red supergiants are
more reddened than the OB stars belonging to the same environment indicating that they are surrounded by local dust material.

If a star becomes a red supergiant at the very beginning of the core He-burning phase, then its red supergiant lifetime will be larger 
than the one of a star which will begin to transform its central helium already in the blue part of the HR diagram and enter
its red supergiant stage in a more advanced evolutionary stage.
Thus the red supergiant lifetimes depend on the way
the star evolves into the red part of the HR diagram after the Main- Sequence phase. This redward evolution depends in turn on the extension of the intermediate convective zone associated with the H-burning shell. Let us recall that
in a convective zone the density gradients are shallower. As a consequence, the larger the intermediate convective zone, the more compact the star and thus
 a blue position in the HR diagram is favored. On the contrary when this intermediate convective zone decreases in mass or even disappears, this favors the inflation of the star and
a rapid evolution into the red part of the HR diagram.  
Mass loss (see e.g.  Meynet 1993) and (rotational) mixing  (Maeder \& Meynet 2001) both reduce the extension of the intermediate convective zone and therefore favor an early entrance into the RSG stage during the core He-burning phase. 

\begin{figure}[h]
\begin{minipage}{8cm}
\centering
\includegraphics[width=8cm, height=8cm]{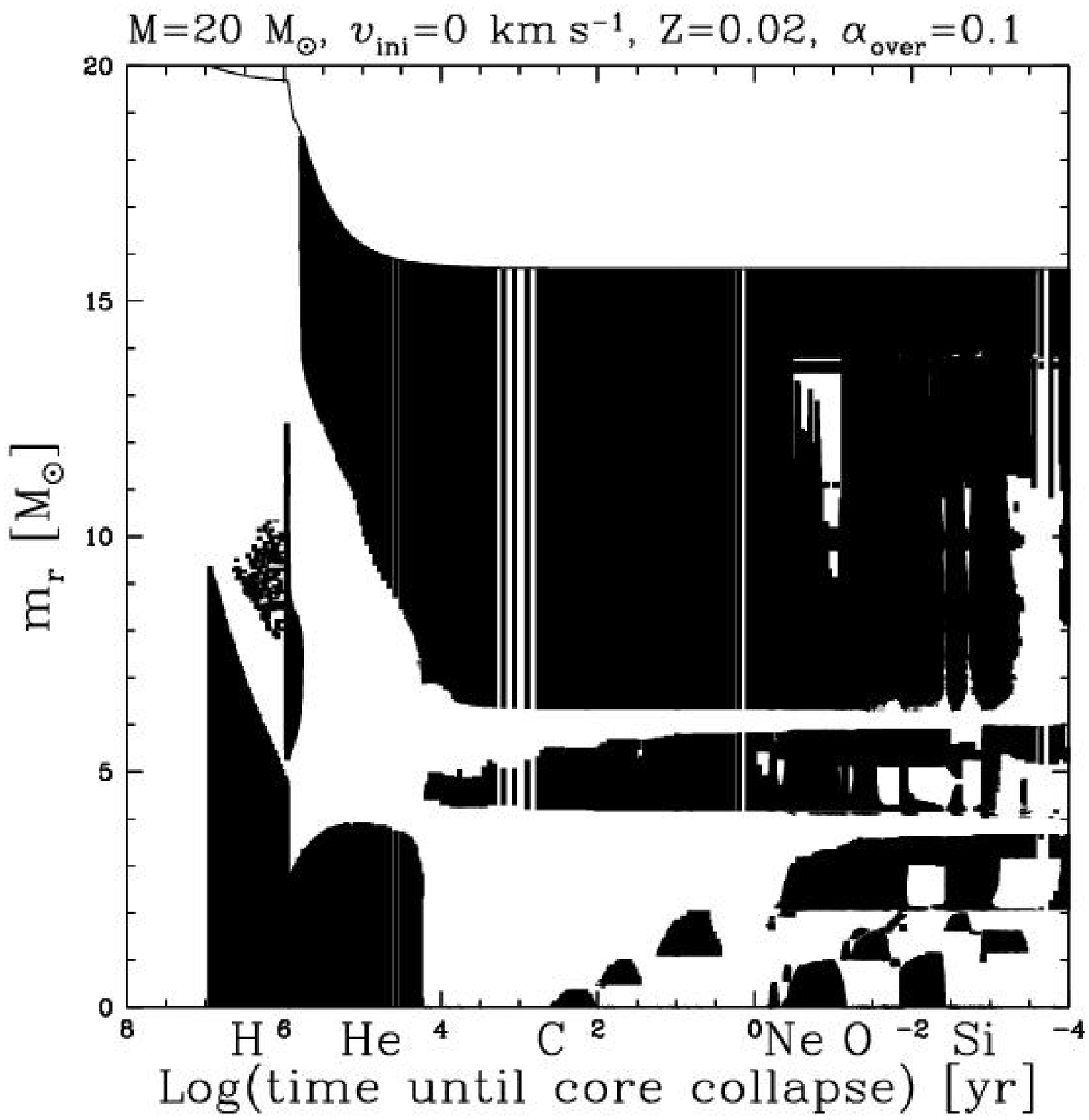}
    \caption{Evolution as a function of the remaining time until core collapse of the total mass and of the
    masses of the convective regions (in black) for a non-rotating solar metallicity 20 M$_\odot$ model. Figure taken from Hirschi et al. (2004).}\label{fig1a}
\vfill
\vfill
\end{minipage}
\hfill
\begin{minipage}{8cm}
\centering
\includegraphics[width=8cm]{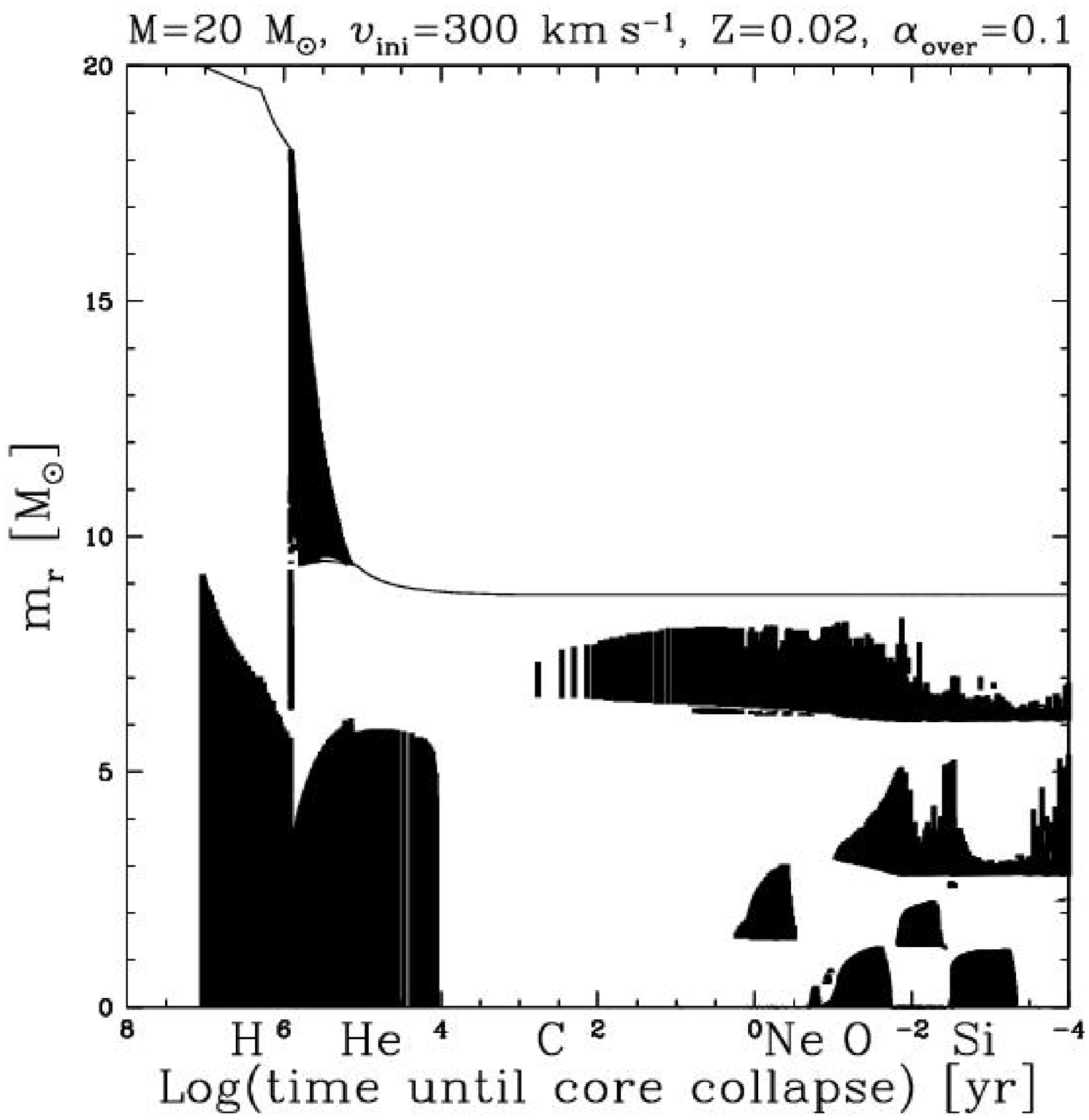}
    \caption{Evolution as a function of the remaining time until core collapse of the total mass and of the
    masses of the convective regions (in black) for a rotating solar metallicity 20 M$_\odot$ model. Figure taken from Hirschi et al. (2004).}\label{fig1b}
\end{minipage}
\end{figure}

\begin{figure}[h]
\begin{minipage}{8cm}
\centering
\includegraphics[width=8cm, height=8cm]{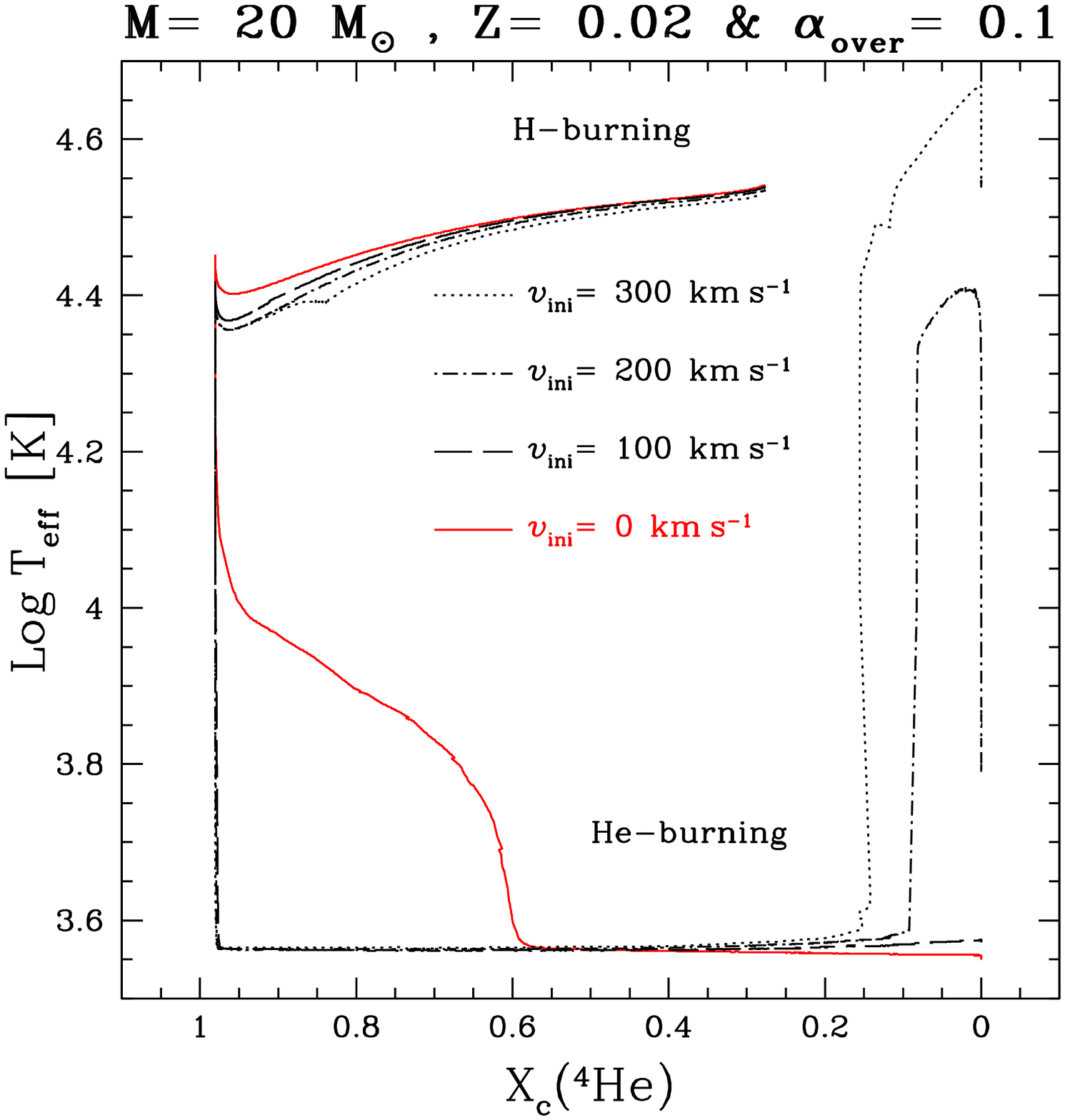}
    \caption{$T_{\rm{eff}}$ vs central helium mass fraction for 20 $M_\odot$ 
models: solid, dashed, dotted-dashed and dotted lines correspond 
respectively to $v_{\rm{ini}}$= 0, 100, 200 and 300 km\,s$^{-1}$. Figure taken from Hirschi et al. (2004)}\label{fig2a}
\vfill
\vfill
\end{minipage}
\hfill
\begin{minipage}{8cm}
\centering
\includegraphics[width=8cm]{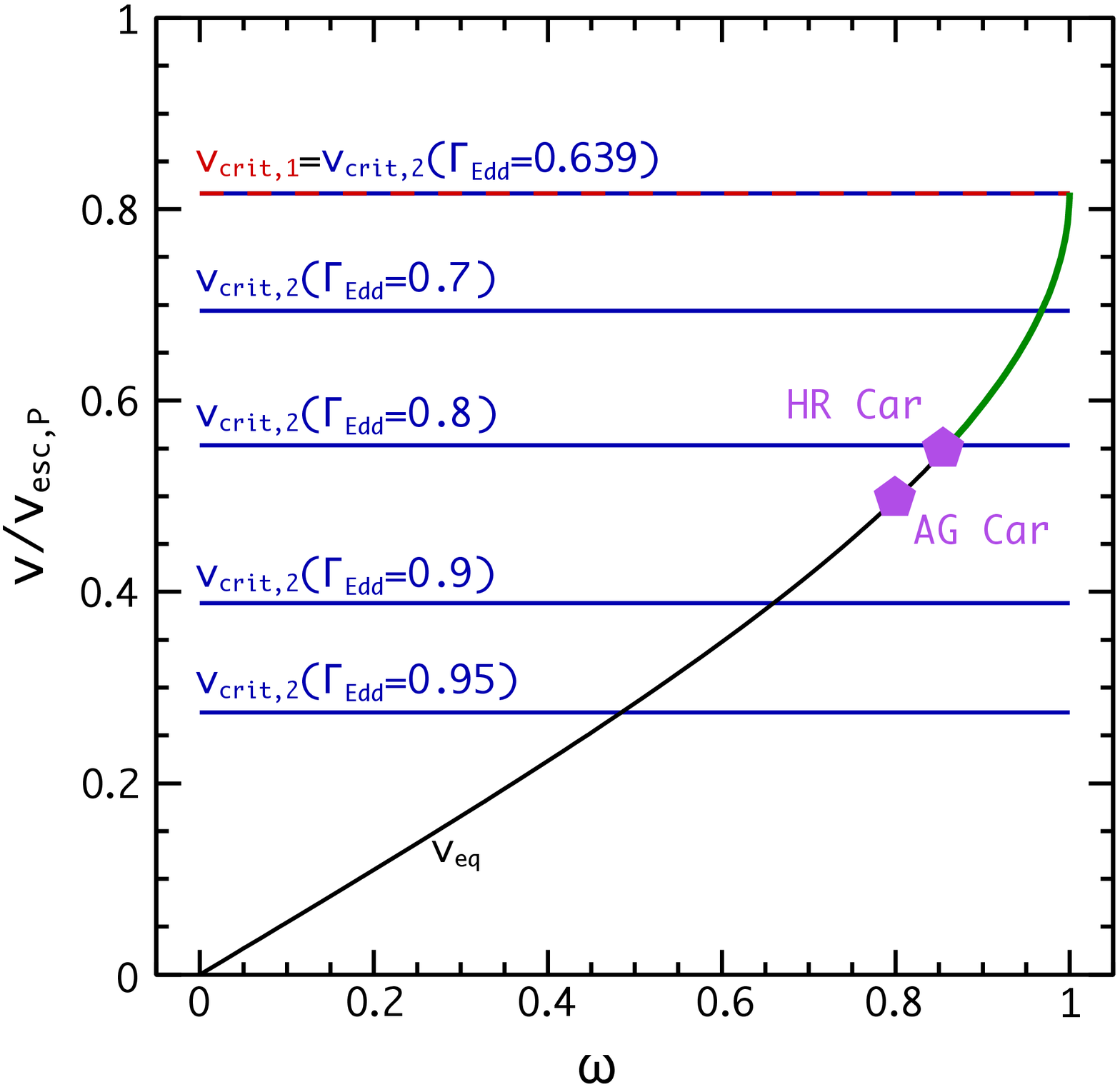}
    \caption{Variation of the linear velocity at the equator as a function of the angular velocity (see text). The horizontal lines
    show the critical velocities for different values of the Eddington factor. Figure taken from Georgy (2010).}\label{fig2b}
\end{minipage}
\end{figure}

Let us illustrate this by looking at Figs.~\ref{fig1a} and \ref{fig1b},  where the structures of  solar metallicity models for two 20 M$_\odot$ stars are shown. 
The models
show the presence of an intermediate convective zone associated with the H-burning shell at the beginning of the core He-burning phase. This intermediate convective zone
stays present for a much longer period in the non-rotating model than in the rotating one (time is in logarithm so it is no so apparent in the figures).
Also the total mass of the non-rotating model  is slightly larger than the total mass of the rotating one at the same stage.
One expects thus that, at the end of the Main-Sequence phase, the non-rotating model will cross the HR gap more slowly. 
This is exactly what happens as can be seen in Fig.~\ref{fig2a}. 

More important but similar effects are observed in metal poor models.
It has been shown by Maeder \& Meynet (2001) that rotation may be the reason for the large number of RSGs observed  in the SMC cluster NGC 330.
Also rotation may be a key ingredient for explaining the B/R ratio estimated in the galaxy Sextans A  from HST imaging by Dohm-Palmer \& Skillman (2002).
They find that if
the non-rotating  stellar models of Schaller et al. (1992) well reproduce the shape of the variation of the B/R ratio as a function of age, the observed ratio is lower than
the model by a factor two.  Rotating models as those presented in Maeder \& Meynet (2001)
would allow to reproduce the observed value.
The link between rotation and red supergiant is also indirectly supported by the observed positive correlation between the population of Be stars, which
are fast rotating B-type star with an equatorial disk and the RSG populations (see Fig. 1b in Meynet et al. 2007).


Once a red supergiant, the star can remain in this state until it explodes in a type IIP core collapse supernova\footnote{Type IIP supernova originates from exploding stars having kept
a massive and extended hydrogen rich envelope.}. But it is not necessarily the case. The star can also
evolve back to the blue part of the HR diagram and end its life as a blue supergiant or even, in some probably rare cases (see below), as 
a Wolf-Rayet star. The important factor governing the blueward evolution from the red supergiant stage is the mass fraction occupied by
the helium core (Giannone 1967). Typically, when the mass fraction of the He-core becomes greater than about 60-70\% of the total actual mass, the star
evolves back to the blue part of the HR diagram. This is what happens in the rotating 20 M$_\odot$ stellar model with $\upsilon_{\rm ini}$= 300 km s$^{-1}$, in which the
core He-burning core occupies more than 65\% of the total mass. The non-rotating model has a core which occupies only about 25\% of the total mass and remains
in the red part of the HR diagram.

As it was the case for the first crossing of the HR gap, this second
crossing  of the HR diagram depends on mass loss (during the red supergiant stage) and mixing (during the previous evolutionary phases). 
Strong mass loss during the red supergiant phase favors a blueward evolution (Salasnich et al. 1999; Vanbeveren et al. 2007; Yoon \& Cantiello 2010)
since it makes the mass fraction of the core larger.  Strong mixing during the previous phases also makes larger cores and
thus favors an evolution from the red to the blue supergiant phase (Hirschi et al. 2004).

Thus we see that the lifetime of the star in the red supergiant stage, as well as the possibility for the star to explode as a type II SN in that stage depend
heavily on mass loss and mixing. 

Let us now recall a few observations which are relevant in that context:
\begin{itemize}
\item the number ratio of blue to red supergiants in clusters with ages between 6.8 and 7.2  (in logarithm of the age in years), or with masses at the turn off between about 12 and 30
M$_\odot$,  increases when the metallicity increases (Meylan \& Maeder 1983; Eggenberger et al. 2002). This observed trend is exactly the opposite
of what is expected from standard grids of models (see the discussion in Langer \& Maeder 1995). 
\item An important number ratio is the relative number of RSGs to WRs.  Massey (2003, see his Fig. 12) shows that this ratio decreases with increasing metallicity, by a factor of 100 over a range of 0.9 dex in metallicity. The strong decrease with the metallicity indicates that the mass range for the progenitors of WRs (RSGs) increases (decreases) when the metallicity increases.
\end{itemize}
These two important features emphasize the importance of metallicity in shaping the blue, red supergiants and Wolf-Rayet populations. Mass loss triggered by radiation increases with the metallicity (see the review by Kudritzki \& Puls 2000) and this is probably one of the main factors responsible for the above observed trends. For instance, when the metallicity increases, stars enter at an earlier phase in the WR phase (making the duration of that phase longer), also stars of lower initial masses can become WR at the end of their lifetime. These two factors are exactly what is required to explain the trend
of the relative number of RSGs and WRs with metallicity. 

For the B/R ratio however this is not the case! Indeed, when the metallicity increases, mass loss rates increase at least during the MS phase and thus  RSGs are favored at high metallicity, which is the contrary of what is observed.
Thus the increase of the mass loss rate with the metallicity, expected for hot stars, cannot be the cause of the observed B/R trend. This
is in fact a counteracting effect. 

Now, we have mentioned that an increase of the mass loss rates {\it during the RSG phase} would shorten the RSGs lifeftime
and lead to the formation of a blue supergiant, or even a WR star in some cases.  At the moment, mass loss rates used in current stellar models
(de Jager et al. 1988) are too weak for producing this kind of evolution for stars in the mass range between 12 and 25-30 M$_\odot$ (at least for standard
non-rotating models). However the uncertainties are large, actually,
the determinations of the mass loss during the RSG phase is still more difficult than in the blue part of the HR diagram due in part to the presence of dust and to various instabilities active in red supergiant atmospheres (e.g. convection becomes supersonic and
turbulent pressure can no longer be ignored). An illustration of the difficulty comes from the determinations of red
supergiant mass loss rates by van Loon et al. (2005). Their study is based on the analysis of optical spectra of a sample of dust-enshrouded red giants in the LMC, complemented with spectroscopic and infrared photometric data from the literature. Comparison with galactic
AGB stars and red supergiants shows excellent agreement for dust-enshrouded objects, but not
for optically bright ones. Dust enshrouded objects show mass loss rates which are
greater by a factor 3-50 than those deduced from optically bright ones! In this context the questions of which stars do
become dust-enshrouded, at which stage, for how long, become critical to make correct prediction of the mass lost by stellar winds. One can also note a very interesting point deduced from the study by  van Loon et al. (2005), that for dust-enshrouded objects mass loss appears to be independent of the metallicity! On the other hand
the formation and duration of the dust enshrouded stage may be metallicity dependent! Thus one sees here that improvements are needed
in order to clarify the situation. Let us just mention that stronger mass loss at high metallicity during the red supergiant phase would go in the right direction for explaining the observed B/R trend with metallicity. Indeed an increase of the mass loss rate during the red supergiant phase at high Z
favors blueward evolution and thus reduces the RSG lifetimes.

We want now to discuss two other indirect arguments possibly supporting the view that red supergiants suffer important mass loss rates. 
The first one concerns an interesting point raised in the review by Smartt (2009). Using archives data, Smartt et al. (2009)
searched for supernova progenitors of type IIP supernovae having occurred in a given volume-limited area of the Universe (28 Mpc) and in a given  time-limited (10.5 yr) interval. They found 20 progenitors for type IIP supernovae. They deduced the mass of the progenitors using
stellar models. They obtained that the mass range of the Type IIP supernova progenitors is comprised between  8.5 (+1/-1.5) M$_\odot$  and 16.5 (+/-1.5) M$_\odot$. 
Why did they not see
progenitors with masses between 18 and 25 M$_\odot$, while red supergiants in this mass range are observed? According to the authors, this lack
is significant. They estimate that the probability of no detection by chance is only 0.018 for a Salpeter IMF.  The authors invoke two reasons for explaining
this lack of massive Type IIP SNe progenitors: 1) the underestimate of the progenitor masses due to dust extinction; 2) the stars in the mass range between 18 and 25 M$_\odot$ do not produce type IIP SNe.  If indeed, stars in the red supergiant stage lose more mass than currently accounted for in models
then it may be that stars evolve back in the blue, ending their life either as blue supergiants, Wolf-Rayet stars or even red supergiants (if they return
back to the red) but with a H-rich envelope too shallow  for producing a type IIP SN.

The second argument is based on the strong helium enrichments observed at the surface of blue supergiants. In a recent work Przybilla et al. (2010 and see the
contribution of Przybilla in the present volume)
determined the He abundance at the surface of solar neighborhood blue supergiants with masses between $\sim$ 9 and 20 M$_\odot$. The enrichments are found to be in the range of 0.32-0.39 (in mass fraction), while rotating models would predict at most values in the range between 0.29 and 0.31 for stars entering the blue supergiant stage directly from the MS. Thus the predicted values are too low. In case the blue supergiants would be actually post red supergiant stars, then He-enrichments
well in line with the spectroscopic values are obtained. Thus, like for the SN argument explained above,  a blueward evolution after the red supergiant phase would overcome the difficulty.

When the star evolves back from the red to the blue, it may encounter instabilities (see e.g. de Jager et al.  2001). Typically when the effective temperature becomes greater than about 7000 K on the blueward track, the main part of the photosphere becomes moderately unstable  (de Jager \& Nieuwenhuijzen 1997).
Observations of yellow supergiants in that region indicate that the approach of this instability region may lead to phases of enhanced mass loss.
These episodes of strong mass loss will help the star to evolve to the blue.

From the discussion above, we see that some interesting constraints would be obtained for stellar models if it would be possible to 
discriminate from observations, blue supergiants originating from MS stars, from blue supergiants originating from red supergiants. As
indicated above helium surface abundances could be a way. But to prove it, we have to show that the He-rich blue supergiants are indeed post red supergiant stars! Is there any possibility, studying the circumstellar environment to find traces of a previous red or even yellow supergiant phase? In these phases
some shell ejections might have occurred which may still be detectable  around at least some blue supergiants.
This remains to be investigated and is probably a very interesting line of research to address both from the theoretical and observational side.
One can wonder also whether the vibrational properties of stars could be used to disentangle these two cases. This is a point which
would also deserve some more theoretical thoughts.

To conclude this section, let us mention another new and interesting problem raised by yellow supergiants.
Yellow supergiants are stars with effective temperatures between 4800 and 7500 K and initial masses above about 9 M$_\odot$. Recently Drout et al. (2009) and Neugent et al. (2010) 
have observed the yellow supergiant populations in M31 and in the SMC. In both cases, they find that rotating and non-rotating models of  Maeder \& Meynet (2001) and of
Meynet \& Maeder (2005) predict too many yellow
supergiants in the mass ranges above 25 M$_\odot$  with respect to the observed number.
It is difficult at the moment to invoke any precise physical cause to this discrepancy. Let us just
make a few remarks: first, since the same discrepancy appears in M31 and in the SMC, this means that metallicity and in particuliar the
metallicity dependance of the mass loss rates probably does not play a role;  2) The models used have no blue loop above 12 M$_\odot$, thus
this discrepancy is not related to a possible blueward evolution after the red supergiant stage. The present models without blue loops already overestimate
the time spent in the yellow supergiant originating directly from the MS phase. In case of a blue loop, the problem would become even more severe.

\section{Luminous Blue Variable stars, transition phase or end point of the evolution?}

The hydrogen surface abundances and the luminosity of the
Luminous Blue Variables indicate that these stars have properties intermediate  between O-type stars and WN stars.
In the evolutionary scenarios recalled at the beginning of this paper, they are interpreted as a transition  stage between these
two types of stars and may be between WNL (here defined as WN stars with hydrogen at the surface) and the WNE stars (interpreted here
as WN stars with no hydrogen at the surface).

Recently however, the light curves of a few supernovae have been found to be consistent with SN ejecta interacting with dense circumstellar material
containing a mass of 10-20 M$_\odot$, {\it i.e.} consistent with the ejected mass during the eruption of LBV stars (type IIn supernovae).  This is the case of SN2006gy and
2006tf (Smith \& McCray 2007; Smith et al. 2008). Also it has been found that the progenitor of SN 2005gl which was a Type IIn is consistent with a very
luminous LBV star, not a RSG (Gal-Yam \& Leonard 2009).
These observations support the view that at least some luminous LBV stars are the end point of the evolution and not a transition phase.
Can such a scenario be explained by single massive star evolution?  
It does not appear to be the case.
LBVs originating from the most massive stars, as for instance $\eta$ Car, lose so much mass in short intervals of time, that these stars will rapidly evolve
away from their LBV stage for becoming a WR stars. Of course such a statement depends on the mass loss history of these stars, on the frequency with
which they undergo strong LBV-type outbursts. Presently, computations modelling the LBV phase by accounting
for an average mass loss rate of about 10$^{-3}$ M$_\odot$ per year during a few tens of  thousands years do not explode as LBV stars.

It may be that strong eruptions are not necessarily linked to LBV stars. 
For instance, Woosley et al. (2007) have invoked the possibility that massive stars could
suffer at the end of their evolution  giant eruptions caused by pulsational pair instability ejections. But this would occur only for initial masses
above 100 M$_\odot$ and according to Smith (2010) these stars are too rare to account for all the observed Type IIn SNe.
Are there other processes producing strong eruptions at the end of massive star evolution? This is of course an important issue which needs
to be addressed in order to make progress on this question.

Let us now come back to one of the main characteristics of LBV stars:  their eruptive nature.
 The physical cause of these eruptions is still subject of discussions. Maeder (1992)
has shown that in the outer layers of these stars, the free fall time becomes longer than the thermal diffusion time. This means that
during the dynamical ejection, the conditions which drive the ejection (producing e.g. the supra-Eddington luminosity) have time to move inside the star 
allowing a significant amount of mass to be ejected (geyser model).  

What are the conditions which are needed for such processes to occur? The anisoptropies of the LBV nebulae support the view that
rotation may play a role. In a fast rotating star, one expects polar ejections and this is quite in line with the peanut shape observed around 
$\eta$ Car, HR Car and AG Car (Groh et al. 2009ab, 2010). 
In order for these anisotropies to be present, the star must rotate at the surface with velocities not too far from the keplerian velocity also called sometimes the classical critical velocity. In the following, the surface angular velocity is noted $\Omega$ and the keplerian velocity $\Omega_{\rm Kep}$.
Typically, $\omega=\Omega/\Omega_{\rm Kep}$ should be greater than 80\%  for the mass flux at the pole be more than 1.65 times the mass flux at the equator (Georgy 2010).
Maeder \& Meynet (2000)  have shown that for luminous stars, i.e. stars sufficiently close to the Eddington limit, the
critical velocity (i.e. a velocity such that the centrifugal acceleration at the equator plus the radiative acceleration  plus the gravity becomes equal to zero)
is no longer given by the classical expression but by a more complex expression involving the
Eddington factor. The higher the Eddington factor, the smaller the critical velocity. In Fig.~\ref{fig2b} the curve shows the relation between the surface angular and the linear
velocity (it would be a straight line for a rigid non-deformed body).
The angular velocity is normalized to the classical critical velocity and the linear velocity to the escape velocity at the pole.
These normalisations and the Roche approximation  (for computing the gravitational potential) make this curve 
independent of the mass, age, metallicity of the star.

The horizontal lines correspond to critical velocities obtained for various Eddington factor $\Gamma=L/L_{\rm Edd}$. As long
as the Eddington factor is inferior to 0.639, the top horizontal line is the critical velocity (note that the critical velocity is lower than the escape velocity. This
reflects the fact that at the critical limit matter is launched in a disk and does not escape to infinity). For Eddington factors below 0.639,  $\omega$
can take any value between 0 and 1. When the Eddington factor increases, the critical velocity decreases and $\omega$ can only take values
between 0 and an upper value defined by the abscissa of the intersection between the horizontal line and the curve. For instance, when 
$\Gamma$ is equal to 0.8, the maximum value of $\omega$ is around 0.85 .  
Using the data given by Groh et al. (2009ab, 2010), we have placed on the red curve the positions of the two LBV stars HR Car and AG Car. These two stars have Eddington factor in the range of 0.8. Thus we see that they rotate very near their critical limit. Interestingly, they still present high enough values of $\omega$ for
allowing significant ansotropies to appear which is in line at least qualitatively with the peanut shape of their nebulae. 
This supports the view that LBV stars may be at the $\Omega\Gamma$-limit as defined in Maeder \& Meynet (2000).

\section{Wolf-Rayet stars and type Ibc supernovae}

\begin{figure}[h]
\begin{minipage}{8cm}
\centering
\includegraphics[width=8cm, height=8cm]{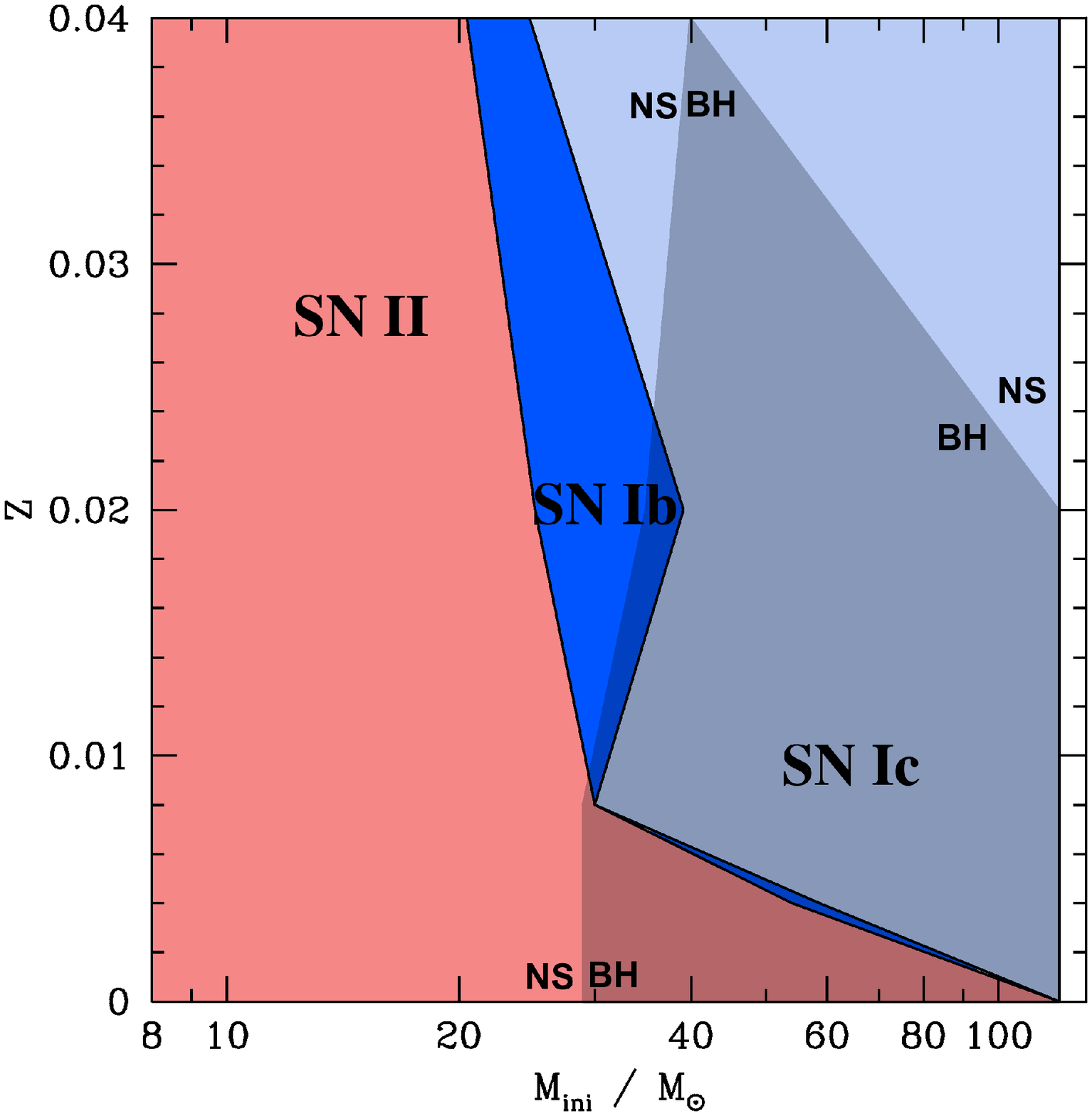}
    \caption{Ranges of masses of different types of SN progenitors at different metallicities. The type of progenitor is indicated in the figure. The shading on the right indicates the area where formation of a black hole (BH) is expected ; elsewhere, the remnant is a neutron star (NS). Figure taken from Georgy et al. (2009).}\label{fig3a}
\vfill
\vfill
\end{minipage}
\hfill
\begin{minipage}{8cm}
\centering
\includegraphics[width=8cm]{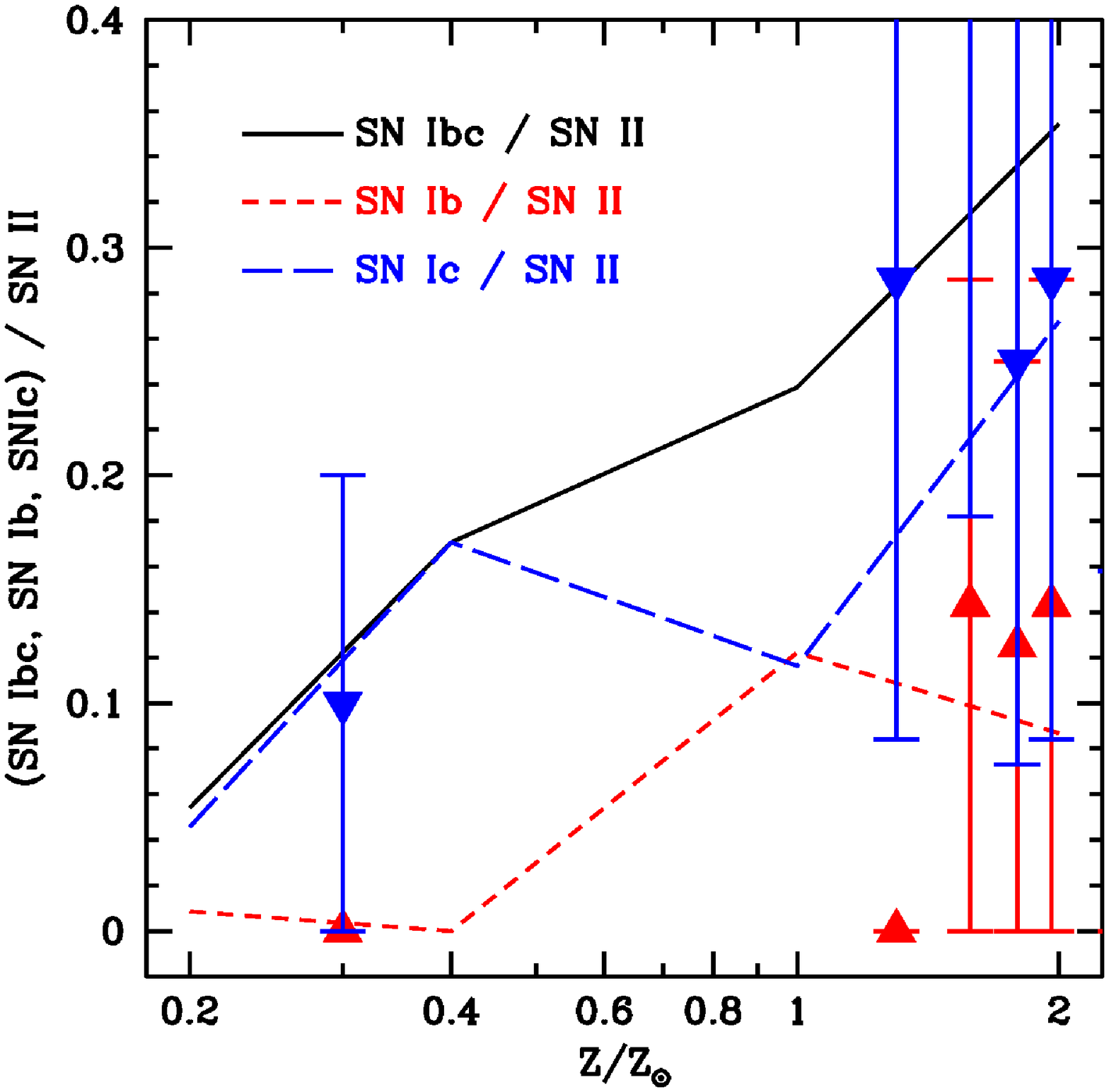}
    \caption{Rates of SN Ic / SN II (blue long--dashed line), SN Ib / SN II (red short--dashed line), and SN Ibc / SN II (black solid line) for the reference case. The points are extracted from the data of Prieto et al. (2008). Figure taken from Georgy et al. (2009).}
    \label{fig3b}
\end{minipage}
\end{figure}

Wolf-Rayet stars are the bare cores of initially massive stars, whose H-rich envelope has been removed by strong stellar winds or through Roche lobe overflow in a close binary system (see the review by Crowther 2007).
Wolf--Rayet stars play a very important role in astrophysics, as signatures
of star formation in galaxies and starbursts, as injectors of chemical elements and of the 
radioactive isotope $^{26}$Al, as  sources of kinetic energy into the interstellar medium and 
as progenitors of supernovae and, likely, as progenitors of long soft $\gamma$--ray bursts.

Rotational  mixing favours the entry into the
WR phase in two ways, firstly
by allowing chemical species produced in the core to diffuse in the radiative envelope and, secondly, by
making the mass of the convective core larger. 
In the non--rotating model,
mass loss by stellar winds is the key physical ingredient which
allows internal chemical products to appear at the surface and thus the  formation of a WR star. The star becomes a WR star
only when sufficiently deep layers are uncovered. In rotating models, the characteristic surface abundances
of WR stars can be obtained through  the effects of mass loss by stellar winds and of
rotational mixing. The action of rotation allow WR stars to appear through the single star scenario even when
the mass loss rates are reduced. To realize that, imagine a star rotating so fast that it would follow a homogenous
evolution. Such a star can become a WR star, i.e. being a star with a log T$_{\rm eff}$ greater than about 4
and a mass fraction of hydrogen at the surface below $\sim$ 0.4 without losing mass!

When models with a time-averaged rotation velocity during the MS phase in the range of the observed
values are considered, then a reasonable number of WR stars can be produced through the single
star scenario even using low mass loss rates (Meynet \& Maeder 2003, 2005).
Also the rotational models well
reproduce the WN/WC ratio at low metallicity and the observed fraction of WR stars in the transition stage WN/WC (phase
during which H- and He-burning products are simultaneously enhanced at the surface).



The number of WC to WN stars  increases with the metallicity (see the review by Massey 2003).
Many attempts have been performed to reproduce this observed trend: for instance the enhanced mass loss rate models of Meynet et al. (1994) provided a good agreement for solar and higher than solar metallicity but produced too few WN stars in metal-poor regions. The inclusion of rotation together with reduced mass loss rates accounting for the effects of clumping improved the situation in the metal poor region, but produced too many WN stars at solar and higher metallicities
(Meynet \& Maeder 2005).
Eldridge \& Vink (2006) show that  models that include the mass-loss metallicity scaling during the WR phase closely reproduce the observed decrease of the relative population of WC over WN stars at low metallicities. However such models severely underestimate the fraction of WR to O-type stars. In that case, to improve the situation, a high proportion of
Wolf-Rayet stars  originating from mass transfer in close binaries should be assumed at all metallicities. For instance at solar metallicity about 75\% of the WR stars should be produced in close binary systems (Eldridge et al. 2008).

The WN/WC number ratio depends also on other factors, in particular on the evolutionary scenario. 
In Meynet \& Maeder (2003), the most massive rotating stars enter into the WR regime already during the MS phase. This feature has good and bad effects. On one hand, it allows these models to well reproduce the variation
of the number fraction of WR to O-type stars since it significantly increases the WR lifetimes. On the other hand, it produces very long WN phases since the star enters into the WR phase having still a huge H-rich envelope. As a consequence,
too low values for the WC/WN ratio are obtained at solar and higher metallicities. 

In Meynet \& Maeder (2003,2005), the hypothesis has been made that 
when a star enters into the WR stage during the MS phase,  it
avoids the Luminous Blue Variable phase. Actually, stars may behave differently. It may well be 
that a star which becomes a WR star during the MS phase, enters a LBV phase after
the core H-burning phase, and then evolves back into the WR regime. 
When this evolutionary scenario is followed, reasonable values for both the WR/O and the WC/WN ratios are obtained.
Indeed the ratios of WR/O and of WC/WN given by these models at the solar metallicity are
0.06 and 0.9 which compare reasonably well with the observed values of 0.1 and 0.9 respectively (Meynet et al. 2008).
Both ratios are not reproduced by the non-rotating models to which a similar scenario is applied.
This discussion illustrates the possible key role that the LBV phase may play in shaping the
WC/WN ratio. 

The nature of the supernova progenitor as a function of the mass and the metallicity can be deduced from stellar models.
This has recently been studied by Georgy et al. (2009) using rotating stellar models (see Fig.~\ref{fig3a}). 
From the presupernova structure, it is also possible to deduce the nature of the supernova explosion and thus to obtain
theoretical estimates for the frequency of the different types of core collapse supernovae as a function of metallicity.
Supposing that any massive star, whether it produces a neutron star or a black hole, produces an observable supernova event,
it is found that single stars can reproduce the variation with the metallicity of the number ratios of type (Ib+Ic)/II, Ib/II and Ic/II supernovae.
Comparisons between observation and theory for the number ratios of Type Ib and Ic supernovae to Type II supernovae are shown in Fig.~\ref{fig3b}.

In case the most massive single stars produce black holes with no supernova events, then single stellar models produce
too few Ibc supernovae at low metallicity and  most Type Ibc
supernovae might then be produced as a result of mass transfer in close binary systems (Pdsiadlowski et al. 1992) . In close binaries, smaller initial mass star 
can still give birth to small mass naked cores. These core would then explode as type Ib or Ic supernova leaving behind a neutron star.
If these naked core are formed near the end of the evolution the star, then their lifetime before explosion could be quite short and
thus the objects may be very rare.

At the moment, it is quite difficult to chose between the single and binary scenario for the progenitors of Type Ibc supernovae, since
both channels can reproduce the observed  variation with the metallicity of the number ratios of Type Ibc to Type II supernovae.
The question of the variation with the metallicity of the number ratios of Type Ib versus Type II and Type Ic versus Type II has not yet been
studied in the framework of the binary scenario.

If the binary scenario is correct, this will decouple the WR populations from the Ibc supernovae. Smartt  (2009)
mentions the fact that 10 supernovae classified as  Ibc have sufficiently deep pre-explosion images available. Curiously, none of them have a
progenitor detected, although, the probability of no detection by chance would be only 11\% (this estimate is obtained by comparing the broad band
magnitudes of known WR stars with the magnitude limits for the progenitors of the 10 Ibc supernovae). 
This probably indicates that the progenitors of the type Ibc comes at least in part from progenitors different from the known WR stars 
and/or that strong shell ejections occurs some time before explosion strongly reducing the magnitude of the progenitor.
As we see, important and interesting points remain to be clarified in order to know the final state of WR stars.

\section{Conclusion}

Let us recall the main three challenges that we mentioned at the beginning of this paper and let us see what we have learned from the present discussion.

\begin{itemize}
\item {\bf the finding of a  consistent explanation of the overall dependence with metallicity of the blue to red supergiant ratio: } rotation appears to play an important role at low metallicity while the interplay of rotation (mainly rotational mixing during the Main-Sequence phase) and mass loss during the red supergiant stage may be the most
important physical mechanisms at solar and higher than solar metallicities. Let us recall here that there is no problem adjusting the mass loss rates or rotation to reproduce the
B/R ratio at a given metallicity. The real challenge is to reproduce the overall trend with a minimum number of hypothesis. 
\item {\bf the driving mechanism for the huge shell ejection undergone by LBV stars:} there are some hints supporting the fact that LBVs are at the $\Omega\Gamma$ limit, i.e.
that the physical conditions for the outbursts to occur involve both the proximity to the Eddington limit and to the critical velocity.  It remains a big challenge for single star models
to make a star explode in the LBV phase. 
\item {\bf the importance of the single and binary channel for explaining the WR and the different core collapse supernova types (II, Ib, Ic) at various metallicities:} the initial mass range of stars becoming WR is different in the single star and the close binary scenario. In the close binary scenario, stars of smaller initial masses can become WR stars.  Typically for instance a 12 M$_\odot$ could become a WR
after having lost its H-rich envelope in a Roche Lobe overflow. This may occur on a timescale allowing
single coeval massive stars to evolve into the RSG stage and thus, one would expect that in the close binary scenario for the formation
of WR stars, we should obtain more frequently the
simultaneous presence of RSGs and WR stars.
The simultaneous presence of WR and RSG in coeval populations are quite rare. Known exceptions are the young clusters of the galactic centre (see e.g. Figer et al. 2004, 2008).
In that case, the high metallicity may explain the simultaneous presence of RSG and WR stars simply because lower initial mass stars can go through the WR stage. Another exception is Westerlund I (Clark et al. 2005). It  is so massive that not all the stars may be coeval. Apart from these cases, RSGs and WR stars are not seen together in coeval populations. 
This may be an indication that formation of WR stars through Roche Lobe overflow from stars
in the mass range between 9 and 25 M$_\odot$ is a relatively rare event. 
\end{itemize}

%
%
%
%
\footnotesize
\beginrefer

\refer Clark, J.~S., Negueruela, I., Crowther, P.~A., \& Goodwin, S.~P.\ 2005, \aap, 434, 949

\refer Crowther, P.~A.\ 2007, \araa, 45, 177         

\refer Dohm-Palmer, R.~C., \& Skillman, E.~D.\ 2002, \aj, 123, 1433

\refer Drout, M.~R., Massey, P., 
Meynet, G., Tokarz, S., \& Caldwell, N.\ 2009, \apj, 703, 441 

\refer Eggenberger, P., Meynet, G., \& Maeder, A.\ 2002, \aap, 386, 576 

\refer Eldridge, J.~J., \& Vink, J.~S.\ 2006, \aap, 452, 295 

\refer Eldridge, J.~J.,  Izzard, R.~G., \& Tout, C.~A.\ 2008, \mnras, 384, 1109

\refer Figer, D.~F., Rich, 
R.~M., Kim, S.~S., Morris, M., \& Serabyn, E.\ 2004, \apj, 601, 319 

\refer Figer, D.~F.\ 2008, 
arXiv:0803.1619

\refer Foellmi, C., Moffat, 
A.~F.~J., \& Guerrero, M.~A.\ 2003a, \mnras, 338, 1025 

\refer Foellmi, C., Moffat, 
A.~F.~J., \& Guerrero, M.~A.\ 2003b, \mnras, 338, 360 

\refer Gal-Yam, A., \& Leonard, D.~C.\ 2009, \nat, 458, 865 

\refer Georgy, C. 2010, PhD thesis, Anisotropic Mass Loss and Stellar Evolution: From Be Stars to Gamma Ray Bursts, Geneva University

\refer Georgy, C., Meynet, G., Walder, R., Folini, D., \& Maeder, A.\ 2009, \aap, 502, 611 

\refer Giannone, P.\ 1967, \zap, 65, 226 

\refer Groh, J.~H., Hillier, 
D.~J., Damineli, A., Whitelock, P.~A., Marang, F., 
\& Rossi, C.\ 2009a, \apj, 698, 1698 

\refer Groh, J.~H., et al.\ 2009b,  \apjl, 705, L25 

\refer Groh, J.~H., Madura, T.~I., Owocki, S.~P., Hillier, D.~J., 
\& Weigelt, G.\ 2010, \apjl, 716, L223 

\refer Hirschi, R., Meynet, G., \& Maeder, A.\ 2004, \aap, 425, 649 

\refer de Jager, C., Lobel, 
A., Nieuwenhuijzen, H., \& Stothers, R.\ 2001, \mnras, 327, 452 

\refer de Jager, C., Nieuwenhuijzen, H., \& van der Hucht, K.~A.\ 1988, \aaps, 72, 259 

\refer de Jager, C., \& Nieuwenhuijzen, H.\ 1997, \mnras, 290, L50 

\refer Kudritzki, R.-P., \& Puls, J.\ 2000, \araa, 38, 613 

\refer Langer, N., \& Maeder, A.\ 1995, \aap, 295, 685 

\refer Levesque, E.~M., 
Massey, P., Olsen, K.~A.~G., Plez, B., Meynet, G., 
\& Maeder, A.\ 2006, \apj, 645, 1102

\refer Levesque, E.~M., 
Massey, P., Olsen, K.~A.~G., Plez, B., Josselin, E., Maeder, A., 
\& Meynet, G.\ 2005, \apj, 628, 973 

\refer Maeder, A.\ 1992, Instabilities 
in Evolved Super- and Hypergiants, Royal Netherlands Academy of Arts and Sciences, p. 138 

\refer Maeder, A., \& Meynet, G.\ 2000, \aap, 361, 159

\refer Maeder, A., \& Meynet, G.\ 2001, \aap, 373, 555

\refer Massey, P.\ 2003, \araa, 41, 15 

\refer Massey, P., Silva, 
D.~R., Levesque, E.~M., Plez, B., Olsen, K.~A.~G., Clayton, G.~C., Meynet, 
G., \& Maeder, A.\ 2009, \apj, 703, 420 

\refer  Meynet, G.\ 1993, The Feedback 
of Chemical Evolution on the Stellar Content of Galaxies, 40  

\refer Meynet, G., \& Maeder, A.\ 2003, \aap, 404, 975 

\refer Meynet, G., \& Maeder, A.\ 2005, \aap, 429, 581 

\refer Meynet, G., Eggenberger, 
P., \& Maeder, A.\ 2007, IAU Symposium, 241, 13

\refer Meynet, G., Maeder, A., Schaller, G., Schaerer, D., \& Charbonnel, C.\ 1994, \aaps, 103, 97

\refer Meynet, G., Ekstr{\"o}m, 
S., Maeder, A., Hirschi, R., Georgy, C., 
\& Beffa, C.\ 2008, IAU Symposium, 250, 147 

\refer Meylan, G., \& Maeder, A.\ 1983, \aap, 124, 84

\refer Neugent, K.~F., Massey, 
P., Skiff, B., Drout, M.~R., Meynet, G., 
\& Olsen, K.~A.~G.\ 2010, \apj, 719, 1784

\refer  Prieto, J.~L., Stanek, 
K.~Z., \& Beacom, J.~F.\ 2008, \apj, 673, 999 

\refer Przybilla, N., Firnstein, M., Nieva, M.~F., Meynet, G., \& Maeder, A.\ 2010, \aap, 517, A38 

\refer Salasnich, B., Bressan, A., \& Chiosi, C.\ 1999, \aap, 342, 131

\refer Schaller, G., Schaerer, D., Meynet, G., \& Maeder, A.\ 1992, \aaps, 96, 269

\refer Smartt, S.~J.\ 2009, \araa, 47, 63 

\refer Smartt, S.~J., Eldridge, 
J.~J., Crockett, R.~M., \& Maund, J.~R.\ 2009, \mnras, 395, 1409 

\refer Smith, N., \& McCray, R.\ 2007, \apjl, 671, L17 

\refer Smith, N., Chornock, R., 
Li, W., Ganeshalingam, M., Silverman, J.~M., Foley, R.~J., Filippenko, 
A.~V., \& Barth, A.~J.\ 2008, \apj, 686, 467

\refer Vanbeveren, D., Van 
Bever, J., \& Belkus, H.\ 2007, \apjl, 662, L107 

\refer van Loon, J.~T., Cioni, M.-R.~L., Zijlstra, A.~A., \& Loup, C.\ 2005, \aap, 438, 273

\refer Yoon, S.-C., \& Cantiello, M.\ 2010, \apjl, 717, L62

\endrefer           
\end{document}